# Chiral Mechanisms Leading to Orbital Quantum Structures in the Nucleon

Dennis Sivers

Portland Physics Institute
4730 SW Macadam #101
Portland, OR. 97239

Spin Physics Center
University of Michigan
Ann Arbor, MI 48109-1120

## Abstract

Color confinement and chiral symmetry specify some important territory for the study of hadronic physics. Any hadron can be defined as a color-singlet composite system of quarks and gluons, the fundamental fields of qcd, while the landscape of the hadronic spectrum is dominated by the fact that two quark flavors, u and d, are characterized by masses small compared to the fundamental scale, $\Lambda_{qcd}$, of this theory. Measurements sensitive to the orbital angular momenta of the SU(3)-color constituents of the nucleon display the interplay of chiral dynamics and confinement in a unique manner. This pageant can be explored by an evaluation, within the context of the Georgi-Manohar chiral quark model, of the normalization of the orbital structure functions $\Delta^N G_{q/p\uparrow}^{front}(x, k_{TN}(x); \mu^2)$ and the normalization of the Boer-Mulders functions $\Delta^N G_{q\uparrow/p}^{front}(x, k_{TN}(x); \mu^2)$ for different quark flavors. The resolution structures in the chiral quark model represent an evaluation of Collins functions $\Delta^N D_{\pi/p\uparrow}(z, k_{TN}; \mu^2)$ for a confined system defined by the quantum numbers of the nucleon in the constituent quark model. The orbital structure functions for antiquarks can also be specified within the basic framework of this approach while the normalization of the gluon orbital structure function requires some additional assumptions.

## I. Introduction

The study of the standard model [1] is complicated by the fact that hadrons consist of color-singlet combinations of the fundamental color-charged objects of quantum chromodynamics, qcd, and by a situation in which no clear theoretical understanding of color confinement has yet been achieved. Lattice regularization of gauge field theory has provided a workable definition of confinement in which the static potential between two massive charged objects is given by

$$V(R) = -\lim_{T \to \infty} \frac{d}{dT} W(R,T) \qquad (1.1)$$

where W(R,T) is a rectangular Wilson loop [2] in Euclidean space-time and in which the potential defined in this way rises linearly with R

$$\lim_{R \to \infty} V(R) = \sigma R \qquad (1.2)$$

The coefficient, $\sigma$, multiplying R is commonly designated the string tension. Computer simulations of lattice qcd have provided compelling evidence that the linear potential, (1.2), is realized with a string tension proportional to the quadratic Casimir of the two color charges.[3] In addition, the simulations show evidence for a "string-like" correction

$$\delta V_s(R) = -\frac{\pi}{12R} \qquad (1.3)$$

to the linear potential.[4] The connection of linear slope (1.2) to Regge trajectories for quark antiquark systems has given an estimate of the corresponding string tension for a massive quark-antiquark pair,

$$\sigma \cong 0.18 GeV^2 \qquad (1.4)$$

The large body of work within lattice qcd on this topic does not provide a formal proof of color confinement but does provide a set of compelling arguments to support the concept. The appearance of a linearly rising static potential can be understood if SU(3) color flux converges to form a "flux-tube" as color sources separate. The "dual-Meissner" model for this phenomenon was proposed by 't Hooft [5] and systematically explored by Mandelstam [6]. This approach has been studied extensively.[7] Alternate theoretical constructions for the mechanism of flux-tube formation involve center-vortices [8] or the AdS-CFT correspondence.[9] A comparative review of these, and other, approaches to the mechanism or mechanisms responsible for color confinement can be found in Ref. 10. Hence, while the concept of color confinement as an exact property of qcd is widely accepted, the diverse theoretical constructions for its implementation preclude any claim that the consequences of confinement can be quantitatively understood at this time. Lattice simulations provide an ever-improving set of approximations for hadronic structure but not all observables are equally appropriate for lattice simulation.

In contrast, the theoretical underpinnings for understanding the constraints of chiral symmetry on hadronic physics are well established. The consequences of PCAC,[11] current algebra[12] and soft pion theorems [13] can be predicted from analytic methods in scattering theory and known symmetry principles.[14] [15] In many instances, the same results can be achieved using techniques involving effective Lagrangians. [16] There now exist numerous instructive summaries [17] that reconcile these methods with the modern understanding of qcd. However, chiral symmetry is only an approximate symmetry of the strong interactions and the comparison of the predictions with data often involves extrapolations or estimates of empirical parameters. Understanding the dynamics of confinement in terms of chiral representations of color sources presents special challenges. The combination of confinement and chiral symmetry therefore presents many complex issues requiring experimental resolution. The rich phenomenology of the hadronic spectrum and of low energy and medium energy hadronic scattering processes offers many opportunities to explore the implications of these two important facets of nonperturbative qcd.

The systematic study of nucleon structure plays a central role in this program. Together, the proton and neutron form the nucleon flavor isospin doublet. Nucleons offer the only stable hadronic targets that can be explored with a number of different probes with controlled kinematics to test nonperturbative qcd. As we shall demonstrate here, a particular aspect of nucleon structure that is sensitive to the interplay of confinement and chiral dynamics involves the orbital angular momentum of the nucleon's color constituents. There are many possible distinct approaches to specifying observables sensitive to parton orbital angular momentum including generalized parton distributions, [18] the polarized gluon asymmetry, [19] and the quark transversity distributions [20]. These observables merit appropriate experimental and theoretical attention. The approach to parton orbital angular momentum followed in this paper involves the measurement of single-spin asymmetries [21]—in particular, the class of single-spin observables that are odd under the symmetry $A_\tau$. For processes involving light quarks, these asymmetries can be shown to be a direct measure of coherent spin-orbit effects in the nucleon. A previous paper [22] presented a simple specification for the normalization of the $A_\tau$-odd orbital structure functions [23] $\Delta^N G_{q/p\uparrow}^{front}(x, k_{TN}(x); \mu^2)$,

$$\int_0^1 dx \Delta^N G_{q/p\uparrow}^{front}(x, k_{TN}(x); \mu^2) = \frac{1}{2}\left\langle L_q^y(\mu^2)\right\rangle = \frac{1}{2}\left\langle \vec{L}_q \cdot \hat{\sigma}_p(\mu^2)\right\rangle \qquad (1.5)$$

and dealt with issues of process dependence in its determination from measurements on single-spin asymmetries. The second expression for the expectation value in (1.5) clarifies the specific definition of the orbital distribution function in that it specifies that the proton polarization is given by a unit vector, $\hat{\sigma}_p$, normal to the scattering plane. This expression for the normalization explicitly displays the connection to spin-orbit dynamics. In this paper we present the analogous, intrinsic geometrical, specification of the Boer-Mulders [24] functions, $\Delta^N G_{q\uparrow/p}^{front}(x, k_{TN}(x); \mu^2)$, describing the spin-orbit dynamics of polarized quarks within an unpolarized proton. Using the Trento conventions [25] the Boer-Mulders functions are thus normalized to

$$\int_0^1 dx \Delta^N G_{q\uparrow/p}^{front}(x, k_{TN}(x); \mu^2) = \frac{1}{2}\left\langle \vec{L}_q \cdot \hat{\sigma}_q(\mu^2) \right\rangle \tag{1.6}$$

in analogy to the normalization of the orbital structure functions. In writing (1.5) and (1.6), we have assumed that the quantities $\left\langle \vec{L}_q \cdot \hat{\sigma}_p(\mu^2) \right\rangle$ and $\left\langle \vec{L}_q \cdot \hat{\sigma}_q(\mu^2) \right\rangle$ are scalars under rotations. They represent currents that can be projected with known techniques. However, we are also interested in normalizing the gluon orbital structure function. This task raises some important issues about confinement. Because the gluon is a gauge boson with only helicity degrees of freedom, the only consistent value for the quantity, $\left\langle \vec{L}_g \cdot \hat{\sigma}_g \right\rangle$, is zero. The only observable involving the angular momentum of the gluon that can transform under rotations as a scalar is $\left\langle \vec{J}_g \cdot \hat{\sigma}_p(\mu^2) \right\rangle$ as specified by, for example, the Ji sum rule. [18] However, in a proton with definite helicity and large momentum along the z-axis, this can be decomposed

$$\left\langle \vec{J}_g \cdot \hat{\sigma}_p(\mu^2) \right\rangle_{\hat{\sigma}_p \cdot \hat{e}_z = 1} = \Delta g(\mu^2) + L_z(\mu^2) \tag{1.7}$$

where $\Delta g(\mu^2)$ is the integral of the gluon's helicity weighted distribution. The orbital angular momentum in this equation can thus be identified by subtraction. With the proton spin transverse to its momentum, there is no gluonic spin operator that can be identified. Nonetheless, there is the possibility of $A_\tau$-odd color currents associated with gluonic fields. Comparison of the Ji sum rule, the $J_z = \frac{1}{2}$ sum rule [26] and the Bakke-Leader-Trueman [27] sum rule therefore specifies that the normalization for the gluon orbital structure function takes the form.

$$\int_0^1 dx \Delta^N G_{g/p\uparrow}^{front}(x, k_{TN}(x); \mu^2) = \frac{1}{2}\left\langle \vec{J}_g \cdot \hat{\sigma}_p(\mu^2) \right\rangle \tag{1.8}$$

To further describe the dynamical information contained in these functions we investigate the combined impact of chiral dynamics and color confinement in a simple version of the chiral quark model first formulated by Manohar and Georgi [28]. This "simple" approach nonetheless demonstrates the surprising complexity of the nucleon's structure. The model allows for the normalization of the orbital structure functions and the Boer-Mulders functions for different flavors of both quarks and antiquarks. We will also show that a consistent description of the rotational invariance of the model combined with the constraints of confinement leads to a lower bound on the value of $\left\langle \vec{J}_g \cdot \hat{\sigma}_p(\mu^2) \right\rangle$ when $\mu^2$ is close to the scale of the chiral transitions.

The remainder of the paper is organized as follows. Sec. II reviews the construction of a local $A_\tau$-odd distribution using the transversity basis. The normalizations of the Boer-Mulders functions are then derived, based on the previously-obtained normalizations for orbital structure functions found in Ref. [22]. Sec. III presents some basic estimates for parton orbital angular momentum using the constraints of color confinement and a simple version of the Georgi-Manohar chiral quark model. The model then leads to a consistent but not necessarily complete description of parton orbital structure. Sec. IV then discusses the results of the model in the context of potential experimental programs and compares the estimates for the distribution functions found here with those of other approaches. We conclude with an appendix with a brief discussion of a Dirac particle bound in a spherically-symmetric potential.

## II. The Construction of $A_\tau$-Odd Parton Distributions for Rotating Constituents.

The conceptual understanding of orbital structure functions is greatly simplified by the existence of a symmetry, $A_\tau$, that can be used to form projection operators for spin-orbit effects. The symmetry can be found by construction. Consider the parity operator, P, acting on a quantum system in the momentum representation,

$$P:\{\vec{p}_i, \vec{\sigma}_i\} \to \{-\vec{p}_i, \vec{\sigma}_i\} \tag{2.1}$$

This form for the parity operator represents the familiar fact that momenta are 3-vectors in space while spins are pseudovectors representing oriented surfaces that do not change sign under a parity transformation. We can then formally define an operator O as a dual form of the parity transformation so that it changes the sign of pseudovectors while preserving vectors.

$$O:\{\vec{p}_i, \vec{\sigma}_i\} \to \{\vec{p}_i, -\vec{\sigma}_i\} \tag{2.2}$$

The operator, $A_\tau$, can then be constructed as a composite, $A_\tau = PO$, giving

$$A_\tau : \{\vec{p}_i, \vec{\sigma}_i\} \rightarrow \{-\vec{p}_i, -\vec{\sigma}_i\}. \tag{2.3}$$

These three operators form a 3-member Abelian group. All observables that can be constructed by measuring only one spin plus some number of momenta in a scattering process are obviously <u>odd</u> under the operator O. Since we can write $O = PA_\tau$ and since all three operators have eigenvalues $\pm 1$, we can classify all single-spin observables into one of two categories:

1. P-odd and $A_\tau$-even.
2. P-even and $A_\tau$-odd.

Within the light-quark sector of the standard model, P-odd single-spin observables are generated by electroweak processes while $A_\tau$-odd single-spin observables are related to dynamic spin-orbit correlations within hadrons. The connection to spin-orbit effects occurs because application of the $A_\tau$ operator defines a local representation of a spin-oriented momentum as required in the formulation of spin-orbit correlations. Note that Eq. (2.3) specifies that the effect of $A_\tau$ on a single isolated particle is the same as that of time reflection. However, $A_\tau$ is <u>not</u> a form of time reversal. It does not impact the causal order of operators and it is not related to CP. The discussion in Ref. [22] explains the distinction more fully and shows how the idempotent projection operators,

$$P_A^\pm = \frac{(1 \pm A_\tau)}{2} \tag{2.4}$$

can be used to project spin-orbit dynamics.

With the aid of the projection operators (2.4) we can specify the number densities for $A_\tau$-odd distributions in terms of the mean orbital angular momentum carried by the constituent involved. The specific construction presented in Ref. [22] can also be applied to normalize the Boer-Mulders functions. For these functions, we consider the local number density averaged over the orientation of proton spin for a quark rotating in the plan normal to its own spin orientation to be:

$$dN_A = P_A^- \{|\Psi_{q\uparrow}|^2 \frac{d^3 p}{E}\} = \frac{2\pi}{\varpi} |\Psi_{q\uparrow}|^2 k_r dk_r dk_y d\phi \tag{2.5}$$

For convenience, the rotation of the transversely polarized quark is chosen to be in the $\hat{x} - \hat{z}$ plane. After the integrations over $dk_r$ and $dk_y$ we have local 4-momentum density in the rest frame of the proton as

$$k_\mu = (\varpi, -k_o \sin\phi, 0, -k_o \cos\phi),$$
$$\varpi = (m_q^2 + \langle k_y^2 \rangle + k_o^2)^{\frac{1}{2}}. \tag{2.6}$$

This gives the light-cone momentum coordinates,

$$k_+ = \omega - k_o \cos\phi$$
$$k_- = \omega + k_o \cos\phi \qquad (2.7)$$
$$k_{TN} = -k_o \sin\phi$$

Using the Trento conventions [25] with the beam particle directed along the $+\hat{z}$ axis, the hard-scattering kinematics are specified by

$$x_o = \omega / M_p$$
$$x' = k_o / M_p \qquad (2.8)$$

and

$$x = k_- / M_p = x_o + x'\cos\phi$$
$$k_{TN}(x) = -M_p [x'^2 - (x - x_o)^2]^{\frac{1}{2}} \qquad (2.9)$$

We find the polarized quark density restricted to front part of the orbit, $\phi \in (0, \pi)$, to be

$$\Delta^N G_{q\uparrow/p}^{front}(x, k_{TN}(x); \mu^2) = -\frac{\langle \vec{L}_q \cdot \hat{\sigma}_q(\mu^2) \rangle}{2\pi} [x'^2 - (x - x_o)^2]^{-\frac{1}{2}} \qquad (2.10)$$

The distribution is therefore normalized so that

$$\int_0^1 dx \Delta^N G_{q\uparrow/p}^{front}(x, k_{TN}(x); \mu^2) = \frac{1}{2} \langle \vec{L}_q \cdot \hat{\sigma}_q(\mu^2) \rangle \qquad (2.11)$$

By symmetry we can define the polarized quark density in the back portion of the target proton to be

$$\Delta^N G_{q\uparrow/p}^{back}(x, k_{TN}(x); \mu^2) = -\Delta^N G_{q\uparrow/p}^{front}(x, k_{TN}(x); \mu^2) \qquad (2.12)$$

Because the constraints of orbit continuity and particle flux conservation restrict the shape in x of the Boer Mulders functions to be the same as that for the orbital distribution functions for quarks and antiquarks, that is

$$\Delta^N G_{q/p\uparrow}^{front}(x, k_{TN}(x); \mu^2) = -\frac{\langle \vec{L}_q \cdot \hat{\sigma}_p(\mu^2) \rangle}{2\pi} [x'^2 - (x - x_o)^2]^{-\frac{1}{2}}$$
$$\Delta^N G_{\bar{q}/p\uparrow}^{front}(x, k_{TN}(x); \mu^2) = -\frac{\langle \vec{L}_{\bar{q}} \cdot \hat{\sigma}_p(\mu^2) \rangle}{2\pi} [x'^2 - (x - x_o)^2]^{-\frac{1}{2}} \qquad (2.13)$$

With an additional assumption concerning the rotational invariance of the appropriate operators we can also normalize the gluon orbital structure function as discussed earlier,

$$\Delta^N G_{g/p\uparrow}^{front}(x, k_{TN}(x); \mu^2) = -\frac{\langle \vec{J}_g \cdot \hat{\sigma}_p(\mu^2) \rangle}{2\pi}[x'^2 - (x-x_o)^2]^{-\frac{1}{2}} \quad (2.14)$$

These functions can all be accurately specified with knowledge of the appropriate expectation value of angular momentum. In each case, the physical observables involving azimuthal asymmetries depend on the impact of initial-state and/or final-state interactions. These interactions generate a process dependence that can be complicated to estimate. The subject of process dependence is discussed further in Ref. [22]. Phenomenological estimates [29] for orbital distributions often assume that the shape of these functions reflect the shape of the corresponding spin-averaged unpolarized distribution. This assumption can make sense in the context of spectator models [30] for orbital distributions that incorporate the full range of initial-state or final-state interactions into the definition of the distribution. The intrinsic geometrical definition presented here is more directly suited to a systematic approach to spin-orbit dynamics such as that found in the Georgi-Manohar chiral quark model. At this point, we want to use this simple model to directly address the dynamics behind the expectation values that appear in (1.5), (1.6), (1.7), (2.10), (2.11), (2.13) and (2.14) so as to complete the specification of these functions.

### III.   Spin-Orbit Dynamics in the Chiral Quark Model

The complexity of proton structure has been a continuing challenge for hadronic physics. A very instructive, minimalist, approach to understanding the inescapable complexity engendered by the combination of confinement and chiral symmetry can be found the chiral quark model proposed by Manohar and Georgi [28]. The model assumes that constituent quarks are composite objects and treats aspects of their structure in terms of interactions with Goldstone bosons. It successfully describes a number of nucleon properties in the resolution scale range 0.2 –1.0 GeV. The model was employed by Eichten, Hinchcliffe and Quigg [31] to explain the flavor asymmetries in the $q\bar{q}$ sea and the experimental observation of a small value for $\Delta\Sigma$ in the context of the $J_z = \frac{1}{2}$ sum rule. It has also been used to describe baryon magnetic moments [32] [33] while the value of this model for systematically understanding parton orbital angular momentum has been emphasized in the work by Cheng and Li [34] and by Song [35]. The application of the model discussed here can be considered a refinement of the work of Song [35] [36] that incorporates further assumptions regarding color confinement and gluon angular momentum.

As the discussion of Sec. II indicates, the existence of $A_\tau$-odd single-spin asymmetries depends on spin-orbit correlations of just the type found in the chiral quark model. The calculations for the normalization of the $A_\tau$-odd effects in this model thus allow us to connect the normalization of these effects to the rich vein of phenomenology mentioned above. In order to specify the orbital distribution functions and the Boer-Mulders functions it is sufficient to use the chiral dynamics of the model to estimate the coefficients $\langle \vec{L}_q \cdot \hat{\sigma}_p \rangle, \langle \vec{L}_q \cdot \hat{\sigma}_q \rangle$, etc. found in Eq's. (1.5),(1.6), (1.7) and (2.10),(2.11) (2.13). The basic model assumptions necessary to calculate these coefficients are:

(1) The valence quark structure for a proton polarized in an arbitrary direction defined as $|P,\uparrow\rangle$ is given by the number densities of constituent quarks; $N(U\uparrow) = 5/3, N(U\downarrow) = 1/3, N(D\uparrow) = 1/3, N(D\downarrow) = 2/3$, where U,D denote the flavors of the composite systems.

(2) A valence constituent quark can experience a virtual chiral fluctuation that displays its color structure or resolves it into a light partonic quark and a pseudoscalar Goldstone boson. The fluctuation necessarily introduces orbital angular momentum and the resolution of the valence quark structure is given by the following transitions.

$$U\uparrow \to [1 - \eta_B - \alpha_c(1 + \varepsilon_s + \varepsilon_o)]u\uparrow ...(L=0)$$
$$+[\eta_B u\downarrow + \alpha_c d\downarrow(\bar{d}u) + \alpha_c\varepsilon_s s\downarrow(\bar{s}u) + \alpha_c\varepsilon_o u\downarrow(\bar{u}u)]...(L=+1) \quad (3.1)$$

$$U\downarrow \to [1 - \eta_B - \alpha_c(1 + \varepsilon_s + \varepsilon_o)]u\downarrow ...(L=0)$$
$$+[\eta_B u\uparrow + \alpha_c d\uparrow(\bar{d}u) + \alpha_c\varepsilon_s s\uparrow(\bar{s}u) + \alpha_c\varepsilon_o u\uparrow(\bar{u}u)]...(L=-1) \quad (3.2)$$

$$D\uparrow \to [1 - \eta_B - \alpha_c(1 + \varepsilon_s + \varepsilon_o)]d\uparrow ...(L=0)$$
$$+[\eta_B d\downarrow + \alpha_c u\downarrow(\bar{u}d) + \alpha_c\varepsilon_s s\downarrow(\bar{s}d) + \alpha_c\varepsilon_o d\downarrow(\bar{d}d)]...(L=+1) \quad (3.3)$$

$$D\downarrow \to [1 - \eta_B - \alpha_c(1 + \varepsilon_s + \varepsilon_o)]d\downarrow ...(L=0)$$
$$+[\eta_B d\uparrow + \alpha_c u\uparrow(\bar{u}d) + \alpha_c\varepsilon_s s\uparrow(\bar{s}d) + \alpha_c\varepsilon_o d\uparrow(\bar{d}d)]...(L=-1) \quad (3.4)$$

In each case, the first line represents an L=0 resolution structure that Preserves the original spin direction of the valence quark while the second line contains the $A_\tau$-odd terms with $\vec{L}\cdot\hat{\sigma}_p = \pm 1$. In all cases the $A_\tau$-odd transitions give $\vec{L}\cdot\hat{\sigma}_q = -1$ where $\hat{\sigma}_q$ refers to the direction of polarization for the "final state" chiral quark. As discussed in Sec. II, these transitions can be considered probability densities.

(3) The transitions represented in (3.1)—(3.4) are assumed to be revealed in the Scale range between 0.2-1.0 GeV and, within the version of this model described in Ref. [35], gluonic degrees of freedom are neglected. However, we will consider the impact of gluonic angular momentum and confinement on the consistency of the model by requiring that the composite system behaves as a $j = \frac{1}{2}$ particle.

The meaning of the different parameters in (3.1)—(3.4) can be understood as follows. The parameter, $\eta_B$, quantifies the amount of $L \neq 0$ found in the wave function for a bound Dirac particle. In the absence of any chiral transitions, the experimental resolution of constituent quark into a parton quark would expose this contribution to orbital angular momentum. For completeness, a brief discussion of the solutions to the bound-state Dirac equation is included in the appendix. The parameter, $\alpha_c$, quantifies the basic transition probability for $U \uparrow \to d \downarrow \pi^+$ while $\alpha_c \varepsilon_s$ gives the rate for the chiral transition $U \uparrow \to s \downarrow K^+$ and $\alpha_c \varepsilon_o$ gives the chiral transition probability for $U \uparrow \to u \downarrow (GB)_o$. We will not make an attempt at this point to do a systematic phenomenological determination of these various model parameters. However, it is instructive to review some of the known constraints. The flavor decomposition of the chiral transitions is determined by the two parameters, $\varepsilon_s$ and $\varepsilon_o$ that specify, respectively the suppression of transitions to kaons and to $\eta$ and $\eta'$. In the full flavor U(3) limit these would be given by $\varepsilon_s = \varepsilon_o = 1$. With maximal SU(3) and U(1) breaking they would be given by $\varepsilon_s = 0$ and $\varepsilon_o = \frac{1}{2}$. A specific model analysis by Song [35] gives an estimate

$$\varepsilon_s + \varepsilon_o \cong 1.0 \tag{3.5}$$

In this section we will give expressions with the full flavor dependence of the model as well as approximate expressions that implement (3.5) in the form $\varepsilon_s \cong \varepsilon_o \cong \frac{1}{2}$. This gives us an opportunity to revisit the issue of flavor dependence within the model later. We will, however, base our numerical estimates on (3.5).

The further specification of parameters requires some discussion of rotational invariance. In any scattering frame, the helicity basis and the transversity basis are related to each other by a boost to the rest frame of the proton, a rotation by ninety degrees, followed by a boost back to the original frame. The separate components of orbital angular momentum and spin do not necessarily transform trivially under the combination of these operations. For example, only the $L_z = 0$ component of a quark of definite helicity rotates into a state of definite transversity and the helicity component of a gluon cannot rotate in this process. Therefore, in order to use (3.1)-(3.4) to normalize orbital distributions we must satisfy ourselves that the expectation values $\langle \vec{L}_q \cdot \hat{\sigma}_p \rangle, \langle \vec{L} \cdot \hat{\sigma}_q \rangle$ and $\langle \vec{J}_g \cdot \hat{\sigma}_p \rangle$ can be treated as scalars under these transformations. To do this we must verify that we can provide a consistent picture of the proton as a whole, within the context of this model, that has the necessary rotational invariance. To do this we will apply the overall constraints of the Ji sum rule [18]

$$\vec{J}_p \cdot \hat{e}_i = \frac{1}{2} = \left\langle \sum_q \vec{J}_q \cdot \hat{\sigma}_p(\mu^2) \right\rangle + \left\langle \vec{J}_g \cdot \hat{\sigma}_p(\mu^2) \right\rangle \tag{3.6}$$

where the quark and gluon components on the RHS can be separately measured by moments of generalized parton distributions. Two separate examples of angular momentum sum rules will be considered. The $J_z = \frac{1}{2}$ sum rule

$$\frac{1}{2} = \frac{1}{2}\Sigma_0 + \Delta g(\mu^2) + \sum_{i=q,g} \left\langle L_i \cdot \hat{e}_z(\mu^2) \right\rangle \tag{3.7}$$

has generated a lot of theoretical attention.[37] In Eq. (3.7) we have chosen a chiral factorization prescription so that $\Sigma_o$ is scale-invariant in the regime where perturbative qcd determines structure evolution. To test consistency of the model, the Bakker-Leader-Trueman sum rule [27]

$$\frac{1}{2} = \frac{1}{2}\Sigma_T(\mu^2) + \sum_{i=q,g} \left\langle \vec{L}_i \cdot \hat{e}_y(\mu^2) \right\rangle \tag{3.8}$$

represents an important additional constraint. All three of these equations thus represent significant independent statements about an isolated composite system with spin. We will examine their impact on calculations with the chiral quark model at the end of this section and use them to estimate values for the parameters $\eta_B$ and $\alpha_c$. At this point we just want to point out that we have initially assumed that it is consistent to neglect corrections $O(\alpha_c^2), O(\eta_B\alpha_c), O(\eta_B^2)$ compared to the leading-order transitions indicated in (3.1)-(3.4). In this section the expectation values calculated with the model parameters will be written without scale indicators until we fix the parameters at approximately the chiral scale by requiring consistency between (3.6)-(3.8).

There is one additional parameter that needs to be specified in order to normalize the orbital angular momentum for individual flavors. In the chiral transitions, we assume that the orbital angular momentum is shared between the chiral quark and the Goldstone boson giving the average values

$$\langle \vec{L}_q \cdot \hat{e} \rangle / \langle \vec{L}_\pi \cdot \hat{e} \rangle = \frac{\lambda_q}{1 - \lambda_q} \tag{3.9}$$

with no flavor dependence and no dependence on scale or orientation. We shall initially fix

$\lambda_q = \frac{1}{4}$ while Song prefers the value $\lambda_q = \frac{1}{3}$. Ultimately, $\lambda_q$ represents an unknown chiral parameter that must be determined experimentally.

It is appropriate to mention at this point that the model transitions given in (3.1)-(3.4) are obviously identical to the dynamical processes that define the Collins functions [38] $\Delta^N D_{\pi/q\uparrow}(z, k_{TN}; \mu^2)$ for the appropriate flavors of quark and boson. The Collins-Heppelman mechanisms [39] provide the opportunity to study these dynamical systems over a wider range of kinematic variables than is possible in a nucleon described in terms of constituent quarks. Measurements of the Collins functions for pseudoscalar bosons in multiple processes can thus directly expand our knowledge of chiral dynamics.

With these parameters defined we can see that the net averaged amount of orbital angular momentum carried by each flavor of quark in this model is given by

$$\langle \vec{L}_u \cdot \hat{\sigma}_p \rangle = \frac{4}{3}\eta_B + \frac{2}{3}\alpha_c(1+\varepsilon_s+\varepsilon_o) - \alpha_c\lambda_q(1+\frac{2}{3}\varepsilon_s - \frac{2}{3}\varepsilon_o) \tag{3.10}$$

$$\langle \vec{L}_d \cdot \hat{\sigma}_p \rangle = -\frac{1}{3}\eta_B - \frac{1}{6}\alpha_c(1+\varepsilon_s+\varepsilon_o) + \alpha_c\lambda_q(\frac{3}{2}+\frac{1}{6}\varepsilon_s - \frac{1}{6}\varepsilon_o), \tag{3.11}$$

$$\langle \vec{L}_s \cdot \hat{\sigma}_p \rangle = \alpha_c\varepsilon_s\lambda_q. \tag{3.12}$$

We can simplify these expressions by choosing $\varepsilon_s \cong \varepsilon_o \cong \frac{1}{2}$ to write

$$\langle \vec{L}_u \cdot \hat{\sigma}_p \rangle \cong \frac{4}{3}(\eta_B + \alpha_c) - \alpha_c\lambda_q \tag{3.13}$$

$$\langle \vec{L}_d \cdot \hat{\sigma}_p \rangle \cong -\frac{1}{3}(\eta_B + \alpha_c) + \frac{3}{2}\alpha_c\lambda_q \tag{3.14}$$

$$\langle \vec{L}_s \cdot \hat{\sigma}_p \rangle \cong \frac{1}{2}\alpha_c\lambda_q \tag{3.15}$$

These expectation values can be inserted directly into (2.13) to normalize the orbital distribution functions for the three quark flavors. Similarly, the net averaged amount of orbital angular momentum carried by each flavor of antiquark in this model is found to be

$$\langle \vec{L}_{\bar{u}} \cdot \hat{\sigma}_p \rangle = (\frac{4}{3}\alpha_c\varepsilon_o - \frac{1}{3}\alpha_c)(\frac{1-\lambda_q}{2}) \tag{3.16}$$

$$\langle \vec{L}_{\bar{d}} \cdot \hat{\sigma}_p \rangle = (-\frac{1}{3}\alpha_c\varepsilon_o + \frac{4}{3}\alpha_c)(\frac{1-\lambda_q}{2}), \tag{3.17}$$

$$\langle \vec{L}_{\bar{s}} \cdot \hat{\sigma}_p \rangle = \alpha_c \varepsilon_s (\frac{1-\lambda_q}{2}). \tag{3.18}$$

These equations can also be simplified by the choice $\varepsilon_s \cong \varepsilon_o \cong \frac{1}{2}$.

$$\langle \vec{L}_{\bar{u}} \cdot \hat{\sigma}_p \rangle \cong \frac{1}{3}\alpha_c (\frac{1-\lambda_q}{2}) \tag{3.19}$$

$$\langle \vec{L}_{\bar{d}} \cdot \hat{\sigma}_p \rangle \cong \frac{7}{6}\alpha_c (\frac{1-\lambda_q}{2}) \tag{3.20}$$

$$\langle \vec{L}_{\bar{s}} \cdot \hat{\sigma}_p \rangle \cong \frac{1}{2}\alpha_c (\frac{1-\lambda_q}{2}) \tag{3.21}$$

These expressions can then be used in (2.13) to normalize the orbital distributions functions for the three flavors of antiquark. Note that in the chiral quark model the contribution of antiquarks to the net orbital angular momentum is not small and the corresponding orbital distribution functions are significant. The chiral quark model gives the expectation values that normalize the Boer- Mulders functions for the three quark flavors as:

$$\langle \vec{L}_u \cdot \hat{\sigma}_u \rangle = -[2\eta_B + (2\alpha_c \varepsilon_o + \alpha_c)\lambda_q] \tag{3.22}$$

$$\langle \vec{L}_d \cdot \hat{\sigma}_d \rangle = -[\eta_B + (\alpha_c \varepsilon_o + 2\alpha_c)\lambda_q] \tag{3.23}$$

$$\langle \vec{L}_s \cdot \hat{\sigma}_s \rangle = -3\alpha_c \varepsilon_s \lambda_q \tag{3.24}$$

Choosing $\varepsilon_s \cong \varepsilon_o \cong \frac{1}{2}$ then gives

$$\langle \vec{L}_u \cdot \hat{\sigma}_u \rangle \cong -(2\eta_B + 2\alpha_c \lambda_q) \tag{3.25}$$

$$\langle \vec{L}_d \cdot \hat{\sigma}_d \rangle \cong -(\eta_B + \frac{5}{2}\alpha_c \lambda_q) \tag{3.26}$$

$$\langle \vec{L}_s \cdot \hat{\sigma}_s \rangle \cong -\frac{3}{2}\alpha_c \lambda_q \tag{3.27}$$

Note that the model gives Boer -Mulders functions that are all negative and that the normalizations are larger than those of the orbital distribution functions of the same flavor. Because the Boer-Mulders functions are related to the Collins functions by crossing relations, this is not a detail of the model but a central feature of the underlying dynamical assumptions embodied in the form of the transitions (3.1)-(3.4). In addition, the antiquarks produced in the virtual fluctuations of this model are embedded in pseudoscalar bosons and, hence, are not polarized. Thus, the Boer –Mulders functions for antiquarks all vanish while the orbital distributions for antiquarks are significant. It is clear that the same underlying chiral dynamics can be responsible for both types of $A_\tau$-odd distribution functions and yet give quite distinct results.

We now turn to the issue of fixing the parameters $\alpha_c, \eta_B$ and $\lambda_c$ in the model. As mentioned earlier, for formulation here is based on the work of Song [35]. That paper chooses the parameters $\varepsilon_o = \varepsilon_s = \frac{1}{2}, \eta_B = 0, \alpha_c = 0.15$ and $\lambda_q = \frac{1}{3}$. It also neglects any angular momentum associated with the gluon but makes it clear that this approximation should not be construed as a prediction of the model as the transitions (3.1) –(3.4) could all take place as specified in the presence of gluon angular momentum. There is, of course, a significant conceptual problem with neglecting the gluons in that the acceleration of color charge by the rotating quarks not bound in Goldstone bosons necessarily drags along gluonic fields. An additional problem can be found by comparing the quark helicity and transversity observables within the context of the model. Choosing $\varepsilon_o = \varepsilon_s = \frac{1}{2}$ for simplicity we see that the total orbital angular momentum carried by quarks and antiquarks in the model is

$$\langle L_q \rangle = \left\langle \sum_{q,\bar{q}} \vec{L}_q \cdot \hat{\sigma}_p \right\rangle = \eta_B + 2\alpha_c \tag{3.28}$$

The transversity distributions are given by the quark fields in the $L = 0$ component of the transitions so that

$$\Sigma_T = \delta^T u + \delta^T d = 1 - \eta_B - 2\alpha_c \tag{3.29}$$

and we have

$$\langle \vec{J}_q \cdot \hat{e}_y \rangle = \frac{1}{2}\Sigma_T + \langle L_q \rangle = \frac{1}{2} + \frac{1}{2}\langle L_q \rangle \tag{3.30}$$

In comparison, the quark helicity distributions calculated in the context of the model give

$$\Sigma_o = \Delta^L u + \Delta^L d + \Delta^L s = 1 - 2\eta_B - 4\alpha_c \tag{3.31}$$

so that

$$\langle \vec{J}_q \cdot \hat{e}_z \rangle = \frac{1}{2}\Sigma_o + \langle L_q \rangle = \frac{1}{2} \tag{3.32}$$

The inconsistency concerning the rotational invariance of the model cannot be resolved without contributions from gluonic angular momentum since rotating color charges necessarily involves rotating color flux. In the presence of gluonic fields we can have scale issues that give some flexibility in defining the quark observables. Consistency of the model and the sum rules (3.6)-(3.8) requires

$$\frac{\langle \vec{J}_g \cdot \hat{\sigma}_p(\mu^2) \rangle}{\langle \vec{J}_q \cdot \hat{\sigma}_p(\mu^2) \rangle} \geq \frac{1}{4} \qquad (3.33)$$

at the scale at which the parameters for the orbital angular momentum expectation values are fixed. We will take a low value for this ratio and give

$$\langle \vec{J}_g \cdot \hat{\sigma}_p(\mu_c^2) \rangle = 0.10 \pm 0.02$$
$$\langle \vec{J}_q \cdot \hat{\sigma}_p(\mu_c^2) \rangle = 0.40 \pm 0.02 \qquad (3.34)$$

where $\mu_c^2 = 1 - -2 GeV^2$ is a scale at which the chiral resolution structures have formed and the $Q^2$- evolution given by perturbative QCD can be considered to begin. At this scale we will choose

$$\Sigma_T(\mu_c^2) \cong \Sigma_o = 0.215 \pm 0.025 \qquad (3.35)$$

where we have assumed that $\Sigma_o$ is specified in a chiral factorization scheme so that its DGLAP evolution has little scale dependence. This allows us to use the orbital expectation values from the model with the assumption that they have scale dependence but no arbitrary angular dependence and we get the constraint on the parameters

$$\eta_B + 2\alpha_c = 0.385 \pm 0.025 \qquad (3.36)$$

at this scale. To determine the ratio between $\eta_B$ and $\alpha_c$ we then use

$$\frac{g_A}{g_B} = \langle 2\vec{J}_q \cdot \hat{\sigma}_p(\mu_c^2) \rangle \frac{5}{3}(1 - \eta_B - \alpha_c) = 1.254 \pm 0.02 \qquad (3.37)$$

As mentioned earlier, the other chiral parameter of the model is chosen to be $\lambda_q = \frac{1}{4}$. This gives the model parameters for calculating the normalization of the orbital structure functions and the Boer Mulders functions to be:

$$\eta_B = 0.052 \pm 0.005$$
$$\alpha_c = 0.120 \pm 0.01 \qquad (3.38)$$

We can compare the normalization using these parameters with what we would have with the original version of the model by Song. We can fix the values for $\langle \vec{L}_q \cdot \hat{\sigma}_p(\mu_c^2) \rangle$ in equations (3.13)-(3.15)

| Orbital functions | Song parameters | This paper |
|---|---|---|
| $u$ quark | 0.150 | $0.197 \pm 0.02$ |
| $d$ quark | 0.025 | $-0.012 \pm 0.01$ |
| $s$ quark | 0.025 | $0.015 \pm 0.005$ |
| Sum of quarks | 0.200 | $0.200 \pm 0.02$ |
|  |  |  |

The normalization of the antiquark orbital distributions is then given at this scale from equarions (3.19)-(2.21)

| Orbital functions | Song parameters | This paper |
|---|---|---|
| $\bar{u}$ antiquark | 0.017 | $0.015 \pm 0.002$ |
| $\bar{d}$ antiquark | 0.058 | $0.053 \pm 0.006$ |
| $\bar{s}$ antiquark | 0.025 | $0.022 \pm 0.002$ |
| Sum of antiquarks | 0.100 | $0.090 \pm 0.01$ |

The values for the mean orbital angular momenta are very similar. The main difference between the models involves objects that Song makes no attempt to calculate but that help us have confidence in the application presented here of normalizing the orbital distributions. The normalization of the Boer-Mulders distributions from eq's. (3.25)-(3.27) at this scale can then be given as

| Boer-Mulders functions | Song parameters | This paper |
|---|---|---|
| $u$ quark | -0.100 | $-0.160 \pm 0.02$ |
| $d$ quark | -0.125 | $-0.125 \pm 0.01$ |
| $s$ quark | -0.075 | $-0.045 \pm 0.004$ |
| Sum of quarks | -0.300 | $-0.330 \pm 0.03$ |

At scales higher that $\mu^2 = 2 GeV^2$ we expect the evolution of these values to be given by perturbative QCD. Our goal here has been to use the Chiral quark model and a small set of assumptions to generate a consistent set for normalizations for the $A_\tau$ - odd distribution functions at this initial scale. The normalization of the gluon orbital distribution

$$\left\langle \vec{J}_g \cdot \hat{\sigma}_p(\mu_c^2) \right\rangle = 0.10 \pm 0.02 \tag{3.39}$$

is not set by the model itself but by indirect arguments concerning consistency. We believe that the total package represents a consistent starting point but at this stage it can be no more than that. The dynamical content of the chiral transitions (3.1)-(3.4) can be described more completely when measurements for the Collins fragmentation functions for pions and kaons from other processes become sufficiently accurate. In addition, we have assumed in the model that resolution structures involving individual constituent quarks are adequate to describe baryon structure at the chiral scale. In doing so, we have neglected resolution structures with diquark quantum numbers. It is not yet clear that this assumption is sufficient to describe baryon structure. At this point, it is appropriate to remind the reader that the transitions (3.1)-(3.4) are identical to the Collins fragmentation functions of the same quantum numbers. They do not necessarily constitute a complete approach to nucleon orbital structure.

This section has presented the first combined normalization of orbital distributions and Boer-Mulders distributions formulated in the transversity basis where the spin-orbit dynamics are calculated in a model originally formulated for other phenomenological purposes. In doing so, we have taken advantage of one of the fundamental features of the transversity formalism in that the normalization of these orbital quantum structures describes intrinsic properties of the proton. The calculations should provide reasonable approximations for these functions at $\mu^2 \cong \mu_c^2$ while at larger momentum scales, the evolution of $\langle \vec{J}_q \cdot \hat{\sigma}_p(\mu^2) \rangle$ and $\langle \vec{J}_g \cdot \hat{\sigma}_p(\mu^2) \rangle$ should track the expectations of perturbative qcd. At some high scale, the ratio of these quantities should approach a constant and it is often assumed

$$\lim_{\mu^2 \to \infty} \frac{\langle \vec{J}_g \cdot \hat{\sigma}_p(\mu^2) \rangle}{\langle \vec{J}_q \cdot \hat{\sigma}_p(\mu^2) \rangle} \square \frac{\langle x_g \rangle}{\langle x_q \rangle} \qquad (3.40)$$

where $\langle x_g \rangle$ and $\langle x_q \rangle$ represent the mean momentum fractions for gluons and quarks. The scale evolution of the Boer-Mulders functions involves additional issues and has not yet been thoroughly investigated. For example, an analogy with the spin tune in storage rings suggests there may be a scale beyond which these functions vanish.

In addition to scale dependence, the ability to test the normalizations calculated here with experimental data involves an understanding of the process dependence of the experimental single-spin asymmetries when expressed in the intrinsic geometrical formalism that is possible in the transversity basis. We will discuss these issues briefly in the next section.

## IV. Discussion and Comparisons

There are numerous experimental programs that have been proposed with the goal to access partonic orbital angular momentum in the nucleon [21] [40]. Among these are efforts to measure deeply virtual Compton scattering and other exclusive hard-scattering processes that can be analyzed in terms of generalized parton distributions [41] [42]. The measurement of the gluon polarization asymmetry [19][43] and of quark transversity distributions [19] [44] can also be shown to provide significant information on orbital angular momentum. In the past few years, it has become widely accepted that measurements on inclusive single-spin asymmetries can be interpreted in terms or orbital structures that involve parton orbital angular momentum [45]. The simple, direct, normalization of these functions for both quarks and antiquarks presented here in the framework of the Georgi-Manohar chiral quark model directly demonstrates the important role that chiral dynamics play in generating these structures. It also explicitly displays the relevant expectation values for orbital angular momentum in the proton that normalize these functions. The strong connection between the dynamical resolution structures in the chiral quark model and the Collins functions, $\Delta^N D_{\pi/q\uparrow}(z, k_{TN}; \mu^2)$ further emphasizes the connection between chiral symmetry and orbital angular momentum. The set of Collins functions and the set of Boer-Mulders functions are necessarily related to each other by generalized crossing relations so that these structures are well specified at the chiral scale by the mechanisms presented here. However, preliminary attempts to understand the dynamics of the polarizing fragmentation functions, $\Delta^N D_{p\uparrow/q}(z, k_{TN}; \mu^2)$, that are, instead, related by crossing to the orbital structure functions suggest that chiral mechanisms with diquark quantum numbers may ultimately be required to explain the dynamics of these functions. The quantity calculated here that is most sensitive to the existence of diquark degrees of freedom in the proton involves the isospin dependence of the expectation values

$$L_q^{I=0}(\mu^2) = \left\langle \sum_{u,d,s} (\vec{L}_q + \vec{L}_{\bar{q}}) \cdot \hat{\sigma}_p(\mu^2) \right\rangle \tag{4.1}$$

and

$$L_q^{I=1}(\mu^2) = \left\langle (\vec{L}_u - \vec{L}_{\bar{u}} - \vec{L}_d + \vec{L}_{\bar{d}}) \cdot \hat{\sigma}_p(\mu^2) \right\rangle \tag{4.2}$$

Diquark spectator models and approximations based on the 1/N approximation imply that

$$L_q^{I=1}(\mu^2) \square\ L_q^{I=0}(\mu^2) \tag{4.3}$$

In addition, two experimental results on inclusive single-spin asymmetries provide preliminary support for the inequality (4.3). First, the approximate mirror symmetry for $pp\uparrow \to \pi^+ X$ and $pp\uparrow \to \pi^- X$ first observed by E-704 [46] and also found at RHIC [47]. Second, the Compass result for small asymmetries in SIDIS on an Isospin zero target (polarized deuteron). [48]

In the calculations based on the chiral quark model we get

$$L_q^{I=1} = 0.268 \pm 0.02$$
$$L_q^{I=0} = 0.303 \pm 0.02$$
(4.4)

and the inequality given in (4.3) is not satisfied at the chiral scale. The qcd-based evolution of these objects should generate a scale dependence, such that

$$\frac{\partial}{\partial \ln \mu^2} L_q^{I=1} \geq 0$$
$$\frac{\partial}{\partial \ln \mu^2} L_q^{I=0} \leq 0$$
(4.5)

as the quark transversities decay and the gluon angular momentum grows with $\mu^2$ and the expense of the isoscalar quark angular momentum. In addition, the process-dependence inherent in the measurement of these $A_\tau$-odd observables could, somehow, preferentially shield isoscalar combinations in specific reactions. A much more plausible conclusion, however, is that orbital quantum fluctuations with diquark quantum numbers must be involved. By including such mechanisms, the Bakke-Leader-Trueman sum rule would still put similar constraints on the isoscalar orbital angular momentum as those described in this paper but the isotriplet combination would be enhanced, perhaps significantly enhanced.

The study of single-spin asymmetries involving quantum structures gains from the multiple formalisms that can be used to describe the process dependence of the experimental results. The most straightforward approach to the inevitable process dependence can be found the twist-3 formalism. [49] [50]. In [51] it has been demonstrated that specific twist-3 calculations can be related to a moment of the orbital distribution function in kinematic regions where both approaches are valid. In [52] it has also been shown that the twist 3 mechanisms display a "universal" pole structure where the color factors and momentum shift engendered by the soft processes in the polarized nucleon can be isolated to give the "response function" of the hard-scattering process to the $A_\tau$-odd mechanisms generating the internal asymmetry. This replicates the factorization properties for orbital distribution functions found in ref [22]. As a consequence, twist-3 calculations can be normalized to the expectation values for orbital angular momentum as indicated in Sec. II on a process-specific term-by-term basis. A similar result seems to be emerging from the gauge link formalism [53]. Because the formalism extends beyond the perturbative approximation, the results appear to be more complicated. The transversity-based formalism presented here offers a direct connection to spin-orbit dynamics. However, much needs to be done on the $\mu^2$ dependence of the intrinsic orbital distributions before quantitative calculations can be considered definitive.

Two recent phenomenological studies offer a direct comparison with our results. Brodsky and Gardner [54] have presented systematic arguments that the small isoscalar signal seen in SIDIS asymmetries for polarized deuterons implies a small gluon orbital distribution. In the chiral quark model, the normalization of the gluon orbital distribution presented here is seen to be small at the chiral scale. However, it is expected to grow at higher factorization scales. The value is constrained by the Bakker-Leader-Trueman sum rule so that a small value of the isoscalar quark orbital distribution would tend to require a larger value for the gluon orbital distribution. Burkardt [55] has expanded his study of models for generalized parton distributions to extract signals sensitive to the Boer-Mulders distributions. His models give Boer-Mulders functions that are large and positive. This conclusion strongly contradicts the chiral quark model results presented here. This may involve a problem with the conventions for the chiral GPDs considered or it may indicate that his analysis has found a source $J = \frac{3}{2}, L = 1$ dynamical mechanisms. This surprising result cannot involve virtual $^3P_0$ pairs or chiral transitions. The independent analysis of Meissner, Metz and Goecke [42] also finds these signals.

**Appendix: Spin-Orbit Effects for bound Dirac Fermions**

Because of charge conjugation, a Dirac particle in a confined system necessarily experiences spin-orbit effects that can lead a nonvanishing expectation value for orbital angular momentum. To see this, we first consider a Dirac particle with an Abelian charge in a central potential. We write the Dirac equation

$$[i\gamma^\mu \partial_\mu - \gamma^0 V + (m+U)]\Psi(x) = 0 \qquad (a.1)$$

in the representation of the Dirac matrices given by

$$\gamma_\mu = \begin{pmatrix} 0 & \sigma_\mu \\ \bar{\sigma}_\mu & 0 \end{pmatrix}, \gamma_5 = \begin{pmatrix} I & 0 \\ 0 & -I \end{pmatrix} \qquad (a.2)$$

with

$$\Psi(\vec{r},t) = \begin{pmatrix} \phi \\ \chi \end{pmatrix} \qquad (a.3)$$

This leads to the coupled equations

$$\vec{\sigma} \cdot \vec{p}\chi + (m+U+V)\phi = E\phi$$
$$\vec{\sigma} \cdot \vec{p}\phi - (m+U-V)\chi = E\chi \qquad (a.4)$$

where $\vec{p}$ is the 3-momentum and E is the energy. Assuming U and V have no angular dependence and making a partial wave expansion, we see that the orbital angular momentum is not a conserved quantum number for the system. Instead, the conserved quantum numbers are found to be $j^2, j_3, K,$ and E with

$$\vec{\Sigma} = \gamma^0 \vec{\gamma} = \begin{pmatrix} \vec{\sigma} & 0 \\ 0 & \vec{\sigma} \end{pmatrix}$$
$$\vec{j} = (\vec{L} + \frac{1}{2}\vec{\Sigma}) \qquad (a.5)$$
$$K = \gamma^0 (\vec{\Sigma} \cdot \vec{L} + 1)$$

The operator, K, quantifies the spin-orbit dynamics that arise because of the correlation between the upper ($\phi$) and lower ($\chi$) components of the Dirac equation in (a.4). The operator equation

$$K\Psi = -\kappa\Psi \qquad (a.6)$$

has the eigenvalues $\kappa = \pm(j+\frac{1}{2})$. There are then found to be two values of orbital angular momentum: $l, l'$, associated with each $\kappa$.

$$\kappa = -(j+\frac{1}{2}) : l = -\kappa+1, l' = -\kappa$$
$$\kappa = +(j+\frac{1}{2}) : l = \kappa, l' = \kappa-1 \qquad (a.7)$$

Note that $l + l' = 2j$. The two orbital values are then distinguished by defining the parity of a solution

$$\gamma^5 \Psi(\vec{r},t) = (-1)^l \Psi(-\vec{r},t) \qquad (a.8)$$

An arbitrary solution [56] of (a.3) with quantum numbers, $j, j_3$ can then be written

$$\Psi_{jj_3}(\vec{r},t) = \begin{pmatrix} f_\kappa(r) A_{jl}^{j_3}(\hat{r}) \\ -ig_\kappa(r) A_{jl'}^{j_3}(\hat{r}) \end{pmatrix}$$

Expressing the Clebsch-Gordon coefficients in the form $(jj'mm'[JM)$ we have

$$\mathsf{A}_{jl}^{j_3} = (l\frac{1}{2}(j_3-\frac{1}{2})\frac{1}{2}[jj_3])Y_l^{j_3-\frac{1}{2}}(\hat{r})\begin{pmatrix}1\\0\end{pmatrix} +$$
$$(l\frac{1}{2}(j_3+\frac{1}{2})\frac{1}{2}[jj_3])Y_l^{j_3+\frac{1}{2}}(\hat{r})\begin{pmatrix}0\\1\end{pmatrix}$$
(a.10)

It is possible to show that

$$\hat{\sigma}\cdot\hat{r}\mathsf{A}_{jl}^{j_3} = -\mathsf{A}_{jl'}^{j_3} \tag{a.11}$$

A more thorough discussion of these solutions is provided by Bhaduri [57]. The $s_{\frac{1}{2}}$ bound state of the system has the quantum numbers $\kappa = -1, j = \frac{1}{2}, l = 0$ and $l' = 1$ and can be written

$$\Psi_{\frac{1}{2}m}(\vec{r}) = \begin{pmatrix} f(r)u_m \\ ig(r)\hat{\sigma}\cdot\hat{r}u_m \end{pmatrix} \tag{a.12}$$

A Dirac particle with non-Abelian charge in a confined system experiences additional spin orbit effects because of the spatial dependence associated with the precession of the nonAbelian charge in group space. This effect was first identified by Wong[58] for an SU(2) gauge theory. Let $I_a$ denote the SU(2) charge and $\xi_\mu(\tau)$ be the world line of a "test" particle. The equations of motion then give

$$\dot{I}_a + g\varepsilon_{abc}A_b^\mu I_c\dot{\xi}_\mu = 0 \tag{a.13}$$

The test particle with thus follow a path in 3-space generated by the precession of $I_a$. A more thorough and elaborate description of this phenomenon can be found in the null-tetrad methods of Carmeli, Charach and Kaye [59]. Specific instructive examples can be found in [60] and [61]. The parameter $\eta_B$ found in Section II gives a simple, phenomenological representation of these dynamical effects that are independent of the underlying chiral transitions in the Georgi-Manohar chiral quark model. In addition, any description of nonAbelian charges must take into account the perturbative Ratcliffe resolution structures [62].

References:

1. See, for example, S. Weinberg, *The Quantum Theory of Fields,* Vols. I, II (Cambridge University Press, New York, 1998)

2. K. Wilson, Phys. Rev. **D10**, 2445 (1974); C. Bachas, Phys Rev. **D33**, 2723 (1986).
3. G. Bali, Phys. Rev. **D62**, 114503 (2000). [ArXiv;hep-lat/0006022]; P. de Forcrand and S. Kratochvila, Nucl. Phys. Proc. Suppl. **119**, 670 [ArXiv:hep-lat/0209004]
4. J. Kuti, [arXiv: hep-lat/0511023] ; M. Luscher and P. Weisz, JHEP **07**, 049 (2002)
5. G.'t Hooft, *High Energy Physics,* (Editorici Compositori, Bologna, 1976)
6. S. Mandelstam, Phys. Reports **23**, 245 (1976)
7. Georges Ripka, *Dual Superconductor Models of Color Confinement,* (Springer-Verlag, Berlin, 2004)
8. G. 't Hooft, Nucl. Physics **B138**, 1 (1978); H. B. Nielsen and P. Olesen, Nucl. Physics **B160**, 380 (1979)
9. J. M. Maldacena, Phys. Rev Lett. **80**, 4859 (1998) [arXiv;hep-th/9803002]
10. J. Greensite, Prog. Particle Nucl. Phys. **51**, 1 (2003)
11. M. Goldberger and S. Treiman, Phys. Rev. **110**, 1178 (1958)
12. S. Adler, Phys. Rev. **140,** 736 (1965); W. Weisberger, Phys. Rev. **143**, 1302 (1966)
13. S. Weinberg, Phys. Rev. Lett. **16**, 879 (1966); ibid. **17**, 616 (1966)
14. J. Goldstone, Nuovo Cimento **19**, 154 (1961)
15. Y. Nambu, Phys. Rev. Lett. **4**, 380 (1960)
16. S. Weinberg, Physica A **96**, 327 (1979)
17. See, for example, *At the Frontier of Particle Physics, The Handbook of QCD,* Vol. I (edited by M. Shifman, World Scientific , Singapore, 2001) H. Leutwyler, "Chiral Dynamics" p. 271; A. Smilga, "Aspects of Chiral Symmetry" p. 317; D. Diakonav and V. Y. Petrov, "Nucleons as Chiral Solitons" p. 359; U. Meissner, "Chiral QCD, Baryon Dynamics" p. 417
18. X. Ji, Phys. Rev. Lett. **78**, 610 (1997)
19. Y. Binder, G. P. Ramsey and D. Sivers (in preparation)
20. Y. Binder, G. P. Ramsey and D. Sivers (in preparation)
21. Proceedings Joint RBRC-UNM Workshop, *Parton Orbital Angular Momentum* 2006, Riken Proceedings Vol. 81
22. D. Sivers, Phys. Rev. **D74**, 094008 (2006), erratum **D75**, e039901 (2007)
23. D. Sivers, Phys. Rev. **D41**, 83 (1990); **D43**, 261 (1991)
24. D. Boer and P. J. Mulders, Phys. Rev. **D57**, 5780 (1994)
25. A. Becchetta, U. D'Alesio, M. Diehl and C.A. Miller, Phys. Rev. **D76**, 117504 (2004)  The conventions in this paper have become standard.
26. J Babcock, E. Monsay, D. Sivers, Phys Rev. **D19**, 1483 (1979)
27. B.L. Bakker,  E. Leader,  T. L. Trueman, Phys. Rev. **D70**, 114001 (2004)

28. A. Manohar and H. Georgi, Nucl Phys. **B234**,189 (1984)
29. M. Anselmino, U. D'Alesio and F. Murgia, Phys. Rev. **D67,** 074010 (2003);**D70**, 074025 (2004); **D71**, 014002 (2005)
30. S. J. Brodsky, D.S. Hwang and I. Schmidt, Phys. Lett. **B530**, 99 (2002); L. Gamberg, G.R. Goldstein and K.A Organessian, Phys. Rev. **D67**, 071504 (2003)
31. E. J. Eichten, I. Hinchliffe and C. Quigg, Phys. Rev. **D45**, 2269 (1992)
32. T.P. Cheng and L.-F. Li, Phys Rev. Lett. **74**, 2872 (1995)
33. X. Song, J.S. McCarthy and H. J. Weber, Phys. Rev. **D55**, 2624 (1997)
34. T.P. Cheng and L.-F. Li, arXiv:hep-ph/970925 (1997)
35. X. Song, arXiv:hep-ph/9801332 (1998)
36. X. Song, arXiv: hep-ph/9712351 (1997)
37. P.G. Ratcliffe, Phys. Lett. **B192**, 180 (1987); X. Song and J. Du, Phys. Rev. **D40**, 2177 (1989); J.W. Qiu, G.P. Ramsey, D. Richards and D. Sivers, Phys. Rev. **D41**, 65 (1990)
38. J.C. Collins, Nucl. Phys. **B396**, 161 (1993)
39. J.C. Collins, S. Heppelman, G. Ladinsky, Nucl. Phys. **B420** 565 (1994)
40. V Barone, A. Drago and P.G. Ratcliffe, Phys. Rep. **359**, 1 (2002)
41. A. V. Rayadushin, arXiv: hep-ph/9704207 (1997); M. Diehl, arXiv:hep-ph/0205208 (2002)
42. S. Meissner, A. Metz, K. Goecke; arXiv:hep-ph/0703176 (2007)
43. M. Stratmann and W. Vogelsang, arXiv:hep-ph/0702083 (2007)
44. G. Afanasev, et. al; arXiv:hep-ph/0703288 (2007)
45. M. Burkardt, Phys. Rev. **D66**, 114005 (2002); **D72**, 094020 (2005)
46. (E-704 Collaberation) D.L. Adams, et al. Phys. Lett. **B264**, 462 (1991)
47. (Brahms Collaberation) F. Videbael, AIP Conf. Proc. **792** ,993 (2005)
48. (Compass Collaberation) V.Y. Alexakhin, et al. Phys. Rev. Lett. **94**, 231 (2005)
49. A.V. Efremov and O.V. Teryaev, Sov. J. Nucl. Phys. **36,** 140 (1982); Phys. Lett. **B150**, 383 (1985)
50. J. Qiu and G. Sterman, Phys. Rev. Lett. **67**, 2264 (1991); Nucl. Phys. **B378**, 52 (1992)
51. X. Ji, J. Qiu, W. Vogelsang and F. Yuan, Phys. Rev. Let. **97** 082002 (2006); Phys. Rev. **D73**, 094017 (2006); Phys. Lett. **B638**, 178 (2006)
52. Y. Koike and K. Tanaka, Phys. Lett. **B646** 232 (2007)
53. A. Bacchetta, C.J. Bomhof, P.J. Mulders, Phys. Rev. **D72**, 034030 (2005); C.J. Bomhof and P.J. Mulders, JHEP 0702, 029 (2007)
54. S.J. Brodsky and S. Gardner, Phys. Lett. **B643,** 206 (2006)
55. M. Burkard, arXiv: hep-ph/0611256 (2006)
56. A.M. Perelomov and V.S. Popov, Sov. J. Nucl. Phys. **14**, 370 (1972)
57. Rajat K. Bhaduri, *Models of the Nucleon*, (Addison Wesley, New York, 1988)
58. S. K. Wong, Nuovo Cim., **65A**, 689 (1970)
59. M. Carmeli, C. Charach, and M. Kaye, Nuovo Cimento **45B** 310 (1978)
60. M. Carmeli, K. Huleihil, Nuovo Cim. **74A**, 245 (1983)
61. D. Sivers, Phys. Rev. **D35**, 707 (1987); Phys. Rev. **D35,** 3231 (1987)
62. P.G. Ratcliffe, Phys. Lett. **B192**, 180 (1987)